\documentclass[conference,letterpaper]{IEEEtran}

\IEEEoverridecommandlockouts

\usepackage{amsmath}
\usepackage{amssymb}
\usepackage{color}
\usepackage{xcolor}
\usepackage{balance}
\usepackage{textcomp}
\usepackage[numbers]{natbib}
\usepackage{url}
\usepackage{tikz}
\usepackage{pgfplotstable,filecontents}
\usetikzlibrary{automata, shapes,arrows, plotmarks, positioning, decorations,calligraphy}
\usepackage{multirow}

\usepackage{pgfplots}
\pgfplotsset{compat=1.14}
\newcommand{\Fig}[1]{Fig.~\ref{#1}}

\newcommand{\todo}[1]{}
\renewcommand{\todo}[1]{{\color{red} TODO: {#1}}}

\usepackage{mathtools}

\usepackage{algorithm}
\usepackage{algpseudocode}
\algnewcommand\algorithmicinput{\textbf{Input:}}
\algnewcommand\INPUT{\item[\algorithmicinput]}
\algnewcommand\algorithmicoutput{\textbf{Output:}}
\algnewcommand\OUTPUT{\item[\algorithmicoutput]}

\newtheorem{theorem}{Theorem}

\newif\ifcomment
\commenttrue

\newcommand{\norm}[1]{\left\| #1 \right\|}
\newcommand{\Pb}[1]{\mathbb{P}\left[#1 \right]}

\def\cn{\mathcal{CN}}

\begin{document}
\title{A Polar Code Based TIN-SIC Scheme for the Unsourced Random Access in the Quasi-Static Fading MAC
\thanks{The research was carried at Skolkovo Institute of Science and Technology and supported by the Russian Science Foundation (project no. 18-19-00673).}
}
\author{
\IEEEauthorblockN{Kirill Andreev, Evgeny Marshakov and Alexey Frolov}
\IEEEauthorblockA{\small Skolkovo Institute of Science and Technology, Moscow, Russia}
{k.andreev@skoltech.ru, evgeny.marshakov@skoltech.ru, al.frolov@skoltech.ru}
}
\maketitle
\begin{abstract}
We consider a problem of unsourced random access in the quasi-static Rayleigh fading channel. In the previous work, the authors have proposed LDPC code based solutions based on joint and treat interference as noise in combination with successive interference cancellation (TIN-SIC) decoder architectures. The authors showed that TIN-SIC decoding significantly outperforms the joint decoding approach and much simpler from the implementation point of view. In this paper, we continue the analysis of TIN-SIC decoding. We derive a finite length achievability bound for TIN-SIC decoder using random coding and propose a practical polar code based TIN-SIC scheme. The latter’s performance becomes significantly better in comparison to LDPC code based solutions and close to the finite length achievability bound.
\end{abstract}
\section{Introduction}
The future of the 5G cellular systems is machine-type communications with a huge number of autonomous devices, short packets, and a lack of centralized coordination. This scenario is actively investigated within the 3GPP standardization committee~\cite{yuan2016non, nikopour2013sparse, 3gpp.R1-164688} and known as massive machine-type communications (mMTC). The main goal for such systems is not the spectral efficiency, but the energy efficiency and connectivity as the majority of autonomous devices are battery-powered.

The previous work on this topic starts from~\cite{polyanskiy2017perspective} where the model of unsourced multiple access was introduced and a finite length random coding bound for the Gaussian multiple access channel (MAC) was derived. The word unsourced means the fact that the users utilize the same encoder or, equivalently, use the same codebook. The improvement of the random coding bound for the Gaussian MAC was given in~\cite{ZPT-isit19}. There is plenty of paper with low-complexity coding schemes for the Gaussian MAC, namely $T$-fold slotted ALOHA (or ALOHA with multi-packet reception) in combination with compute-and-forward strategy \cite{ordentlich2017low, facenda2019efficient}, $T$-fold irregular repetition slotted ALOHA (IRSA) in combination with LDPC codes~\cite{vem2017user, 10.1007/978-3-030-01168-0_15, Glebov2019}, $T$-fold IRSA in combination with polar codes~\cite{Marshakov2019Polar}, coupled compressive sensing~\cite{amalladinne2018coupled, calderbank2018chirrup}, sparse regression codes~\cite{Fengler2019sparcs}, sparse spreading~\cite{pradhan2019joint}, polar codes with random spreading and correlation-based energy detector~\cite{pradhan2019polar}.

The Gaussian MAC is an idealized channel model. The synchronous quasi-static fading MAC has been considered in~\cite{kowshik2019quasi, FadingFull2019, FadingISIT2019, kowshik2019fundamental}, where achievability bounds and LDPC code based practical schemes have been proposed. The asynchronous quasi-static fading MAC has been considered in \cite{FadingASYNC2019, Andreev2019Asilomar, amalladinne2019async, chen2017sparse}. Bounds and solutions for the MIMO MAC have been proposed in~\cite{fengler2019massive}. Actually, the authors assume multiple antennas at the receiver only which is reasonable to reduce the energy consumption at transmitters. We also note that the mentioned above papers assume the absence of channel state information (no-CSI assumption) both at transmitters and the receiver. The reason is as follows: it is extremely difficult to estimate the channel for a huge number of devices transmitting short packets. 

In this paper, we focus on the synchronous quasi-static fading MAC. In \cite{FadingASYNC2019, Andreev2019Asilomar} an LDPC code based TIN-SIC decoder architecture was proposed. The main idea can be explained as follows: at each step, we decode the strongest user codeword (TIN part) and remove it from the channel output (SIC part). SIC part is of the most interest as the fading coefficient is unknown. We find a residual channel output $Y'$ as $Y - \mathrm{P}_{<\mathcal{C}_0>} Y$, where $Y$ is the received vector and $\mathrm{P}_{<\mathcal{C}_0>} Y$ is an orthogonal projection of $Y$ onto space spanned by the set $\mathcal{C}_0$ of already decoded codewords. It was shown that TIN-SIC decoding significantly outperforms the joint decoding approach and much simpler from the implementation point of view.

We continue the analysis of TIN-SIC decoding. We derive a finite length achievability bound for TIN-SIC decoder using random coding and show it to be better in comparison to the bound from~\cite{FadingFull2019, FadingISIT2019}. We also improve the practical scheme by utilizing polar codes~\cite{Arikan} in it, taking into account the excellent decoding performance of polar codes for short blocklength. For this scheme to work we need the coarse fading coefficient estimate which is done by means of Expectation-Maximization (EM) clustering. We note that in contrast to~\cite{Marshakov2019Polar} where a polar code based IRSA scheme was proposed for the Gaussian MAC here we utilize carefully constructed single user polar codes. The performance of the resulting scheme is found to be significantly better in comparison to LDPC code based solutions and close to the finite length achievability bound.
\section{System model}
\subsection{Rayleigh fading channel}
Consider the typical mMTC scenario with a potentially unbounded number of users $K_\text{tot}$ in the system with only $K_a \ll K_\text{tot}$ of them being active at each time instant. Communication proceeds in a frame-synchronized fashion with the frame length equals to $n$. Each user has $k$ bits to transmit within a frame. Within this paper we consider single antenna scenario and assume the presence of an ideal synchronization. All users utilize the same message set $[M] \triangleq \{1, \ldots, M\}$ and the same codebook $\mathcal{C} = \{X^n(W)\}_{W=1}^M$. In order to send a message $W_i$ the $i$-th user will use a codeword $X_i = X^n(W_i)$. Every transmission has a power constraint $\norm{X^n_i}^2 \leq nP$. The signal received within a frame is
\begin{equation}
\label{eq:channel}
Y^n=\sum_{i=1}^{K_a} X_i^n\cdot \mathrm{diag}(H^n_i)+Z^n,
\end{equation}
where $Z^n \sim\mathcal{CN}\left(0,I_n\right)$ are i.i.d. realizations of noise and $H^n_i$ are the fading coefficients which are independent of $X^n_i$ and $Z^n$. Recall that we assume $H^n_i$ to be unknown both at the transmitters and the receiver (no-CSI assumption).  We use slow fading channel model and assume the fading coefficients to be constant during $n_1< n$ channel uses (so-called quasi-static property or channel coherence time).
\subsection{Per-user probability of error}
Due to the same codebook assumption we require the decoder to return only the list of transmitted messages (up to permutation) and use the per-user probability of error (PUPE)~\cite{polyanskiy2017perspective} as a performance measure. PUPE is defined as follows.
\begin{equation}
\label{eq:def1}
P_{e}=\frac{1}{K_a}\sum_{j=1}^{K_a}\Pb{E_j},
\end{equation}
where $E_j\triangleq\{W_j \notin L(Y^n)\}\cup \{W_j= W_i \text{ for some } i \neq j\}$, $Y^n$ is the channel output~\eqref{eq:channel} and $L(Y^n)$ is the list of messages returned by the decoder.

We measure the energy efficiency with use of energy per information bit $E_b/N_0 = nP/k$ required to achieve $P_e \leq \varepsilon$. 
\subsection{$T$-fold ALOHA}
The $T$-fold ALOHA scheme is a good candidate for a practical solution because it reduces the total number of simultaneous transmissions by splitting the frame into slots. Let $T,n_1 \in \mathbb{N}$ such that $T<K_a$ and $n_1<n$. The frame of length $n$ is partitioned into $V=n/n_1$ slots of length $n_1$. The common codebook is of blocklength $n_1$ and thus may use a larger power $VP$ per degree of freedom. Each user chooses a slot to send his message uniformly at random independently of other users. If there are $r$ users placing their codewords in a particular slot, then the signal received in a slot is as follows
\begin{equation}\label{eq:ch_n1}
Y^{n_1} = \sum_{i=1}^r H_i X_i^{n_1} + Z^{n_1}\,, \qquad W_i, \stackrel{iid}{\sim}\mathrm{Unif}[M]\,.
\end{equation}
and $H_i\stackrel{iid}{\sim}\cn(0,1)$, $i=1, \ldots, r$ (in accordance with Rayleigh block-fadin channel model).

The term ``$T$-fold'' means that the decoder aims to resolve the collisions of order up to $T$, i.e. given some value of $r$, the decoder can estimate all $r \le T$ messages with good reliability, while if $r>T$ users were transmitting then no guarantees can be given. The value of $T$ controls the overall decoder complexity. The case $T=1$ corresponds to the usual slotted ALOHA.

In what follows we do not make the long channel coherence time assumption, i.e. the quasi-static fading property applies only to a slot rather than to the whole frame. Thus, the application of IRSA scheme~\cite{liva2011graph} becomes impossible as fading coefficients change at random between slots.
\subsection{TIN-SIC decoder}
For now, consider the slot decoding by means of aforementioned TIN-SIC decoder.
As soon as a single codeword decoding algorithm is an ordinary decoder (LDPC or polar), the main part of the TIN-SIC algorithm is the known codeword subtraction. Let us consider this process in more detail.

Let us denote the TIN decoder by $\mathbb{D}_{TIN}$. It returns some codeword or an empty set (or failure). The task of the decoder is to return the set of unique codewords extracted from the signal mixture $Y$ (see~\eqref{eq:ch_n1}). We omit the subscript $n_1$ here and in what follows. 

Let the $\mathcal{C}_0 = \{X_1, X_2, \ldots, X_\ell\}$, be the set of successfully decoded codewords. The SIC procedure needs to perform the subtraction. We find a residual channel output $Y'$ as 
\[
Y' = Y - \mathrm{P}_{<\mathcal{C}_0>} Y,
\]
where $\mathrm{P}_{<\mathcal{C}_0>} Y$ is an orthogonal projection of $Y$ onto the space spanned by the set of already decoded codewords $\mathcal{C}_0$. Note that the projection can be calculated as (we emphasize that we also estimate CSI as an intermediate step)
\begin{equation}
\left(\hat{H}_1, \ldots, \hat{H}_\ell\right) = 
\mathop{\arg\min}_{({H}_1, \ldots, {H}_\ell)}{\left\|Y - \sum\limits_{l=1}^{\ell} H_l X_l\right\|^2}
\label{eq:cancellation}
\end{equation}
and thus
\[
Y' = Y - \sum\limits_{l=1}^{\ell} \hat{H}_l X_l.
\]

This is algorithm is inspired by a well-known orthogonal matching pursuit (OMP) approach described in~\cite{Cai2011}. Note that MMSE-based analog~\cite{Sparrer2016} is not required here because the number of channel uses is high enough.
\begin{algorithm}
\caption{TIN-SIC decoder\label{alg:SIC}}
\begin{algorithmic}
\State $\mathcal{C}_0 \gets \varnothing$
\State $Y' \gets Y$
\For {$i = 1, \ldots, T$} \Comment Run $T$ decoding attempts
\State $X_i = \mathbb{D}_{TIN}(Y')$ \Comment Perform a TIN decoding attempt
\If{$X_i = \emptyset$}
\State{Break}
\EndIf  
\State $\mathcal{C}_0 = \mathcal{C}_0 \cup X_i$ \Comment Update the set of unique codewords
\State $\ell \gets |\mathcal{C}_0|$
\State Calculate $(\hat{H}_1, \ldots, \hat{H}_\ell)$ \Comment See~\eqref{eq:cancellation}
\State $Y' = Y - \sum\limits_{l=1}^{\ell} \hat{H}_l  X_l$ \Comment Perform cancellation
\EndFor
\end{algorithmic}
\end{algorithm}
The formal algorithm description is given by Algorithm~\ref{alg:SIC}.
\section{TIN-SIC achievability bound}
In this section, we discuss our main achievability bound for $T$--fold ALOHA protocol in combination with TIN-SIC slot decoder. Let us fix some slot code $\mathcal{C}$ and assume that the slot decoder is aware of the actual number of users transmitting in a slot (genie assumption). Then the PUPE per slot can be calculated as follows
\[
p_{e,\text{genie}}(\mathcal{C}, T, r) = {\frac{1}{r}} \sum_{i=1}^r \Pb{W_i \not\in L(Y, T, r)}.
\]
and the overall PUPE of the $T$-fold ALOHA access scheme is bounded by
\begin{eqnarray*}
\varepsilon_{T, \text{genie}}(\mathcal{C}) &\triangleq& 1-\sum_{r=1}^{K_a}(1-p_{e,\text{genie}}(\mathcal{C}, T, r)) \\
&\times& \binom{K_a-1}{r-1}\left(\frac{1}{V}\right)^{r-1}\left(1-\frac{1}{V}\right)^{K_a-r}.
\end{eqnarray*}

Our main contribution in this section is a random coding bound for $p_{e,\text{genie}}(\mathcal{C}, T, r)$. Before we shift to the theorem statement and the proof, let us discuss the genie assumptions which we use in what follows:
\begin{itemize}
    \item {\bf Assumption 1.} The exact number of users in a slot is known to the decoder. Actually, we do not need this information in a case $r \geq T$. Indeed, we will decode $T$ strongest users and stop. But in a case $r < T$ and the use of maximum likelihood (ML) TIN decoder we will find $T-r$ false messages. 
    \item {\bf Assumption 2.} We assume perfect interference cancellation, i.e. as soon as the codeword is found the decoder is given the exact value of the fading realization.
\end{itemize}
Due to genie assumptions, the proposed below bound is not a true achievability bound but as we will see later, the practical scheme is rather close to it. At the same time, we note that we do not use genie assumptions in the practical scheme.

We use random coding with \textbf{Gaussian ensemble} $\mathcal{E} (M, n_1)$: $X(W_i) \mathop{\sim}\limits^{iid} \cn(0,P'I_{n_1})$ where $P'\leq P$. If $\norm{X(W_i)}^2>n_1P$ then that user sends $0$.

The last thing we need to specify is the TIN decoding method. Here we follow~\cite{Polyanskiy2013Fading} and use a projection decoder
\[
\hat{X}=\arg\max_{X \in \mathcal{C}}\norm{\mathrm{P}_{X}Y}^2.
\]

The paper~\cite{Polyanskiy2013Fading} gives an achievability bound $R^*_\text{noCSI}(n, \varepsilon, P, H)$ for the code rate when a projection decoder is applied for a single user channel $Y = H X + Z$, $Z \sim \mathop{\sim}\limits^{iid} \cn(0, I_n)$, the required error probability is $\varepsilon$ and $H$ is a fading coefficient which can have arbitrarily pdf. It is better for us to work with a bound on the error probability thus we introduce the function
\[
p^*(M,n,P, H) = \inf_{\varepsilon} \left\{\varepsilon: \frac{\log M}{n} \leq R^*_\text{noCSI}(n, \varepsilon, P, H) \right\}.
\]

\begin{theorem}
Let $P' < P$ be fixed. Under Assumptions $1$ and $2$, there exists a code $\mathcal{C} \in \mathcal{E} (M, n_1)$ such that for TIN-SIC decoder the following bound holds
\begin{flalign*}
&p_{e,\text{genie}}(\mathcal{C}, T, r) \leq p_0 + \max\left\{\frac{r-T}{r}, 0\right\}\\
& + \frac{1}{r} \sum\limits_{i=1}^{\min\{r,T\}} (r-i+1) \mathbb{E}\left[ p^*\left(M,n_1,\frac{P' V}{1 + \sum_{j=i+1}^r |H_j|^2}, H_i\right)\right],
\end{flalign*}
where 
$$
p_0 =\frac{\binom{K_a}{2}}{M}+K_a\Pb{\frac{P'}{2}\sum_{i=1}^{2n_1}S_i^2>nP}, \quad S_i \mathop{\sim}\limits^{iid} \mathcal{N}(0,1)
$$
and the expectation is taken over $H_1, H_2, \ldots, H_r$: $|H_1| \geq |H_2| \geq \ldots \geq |H_r|$.
\end{theorem}

\begin{IEEEproof}
Without loss of generality, let us assume that $|H_1| \geq |H_2| \geq \ldots \geq |H_r|$ and the corresponding users' codewords are $X_1, X_2, \ldots, X_r$. Consider the TIN-SIC decoder. In case of error at the $i$-th step we have PUPE = $(r-i+1)/r$ and thus
\begin{flalign}
&p_{e,\text{genie}}(\mathcal{C}, T, r) \leq \max\left\{\frac{r-T}{r}, 0\right\} \nonumber\\
&+\frac{1}{r} \sum\limits_{i=1}^{\min\{r,T\}} \left( (r-i+1) \Pb{X_i \ne \hat{X_i} | \left\{X_j\right\}_{j=1} ^{i-1}}\right),
\label{eq:pupe_bound}
\end{flalign}
where $\hat{X_i} = \mathbb{D}_{TIN}(Y')$ and $Y'$ is a residual channel output at step $i$.

Let us calculate the average PUPE over the ensemble $\mathcal{E} (M, n_1)$, i.e. $\mathbb{E}_{X_1, \ldots, X_r}\left[p_{e,\text{genie}}(\mathcal{C}, T, r)\right]$. First, by $p_0$ we upper bound the probabilities of the events $E_1 = \{W_j= W_i \text{ for some } i \neq j, \:\: i,j \in [K_a]\}$ and $E_2 = \{\norm{X(W_i))} > n_1P \text{ for some } i \in [K_a]\}$.  

Consider the $i$-th step, let us estimate
$$
\mathbb{E}_{X_1, \ldots, X_r}\left[\Pb{X_i \ne \hat{X_i} | \left\{X_j\right\}_{j=1} ^{i-1}}\right].
$$

Clearly, for the TIN-SIC algorithm, we can consider the equivalent single user channel model
\begin{equation}
\label{eq:tin_interference}
Y' = X_iH_i + \sum\limits_{j=i + 1}^{r}X_j H_j + \sum\limits_{j=1}^{i - 1}X_j \left(H_j - \hat{H}_j\right) + Z,
\end{equation}
where the first term is the signal to be decoded, while the last three terms are the interference, interference caused by non-ideal SIC procedure and noise respectively. $\hat{H}$ is the CSI estimate, see Algorithm~\ref{alg:SIC}.

Due to Assumption~$2$ we have
\[
\sum\limits_{j=1}^{i - 1}X_j \left(H_j - \hat{H}_j\right) = 0.
\]

We are going to apply the bound $p^*(M,n,P, H)$, the only problem is caused by the ordered statistics $H_i$. But at the same time if we fix $H_1, H_2, \ldots, H_r$, we can use the bound $p^*\left(M,n_1,\frac{P' V}{1 + \sum_{j=i+1}^r |H_j|^2}, H_i\right)$. To finish the proof we calculate the expectation over $H_1, H_2, \ldots, H_r$. 


\end{IEEEproof}
\section{Polar code based scheme}
\subsection{TIN decoding}
\begin{figure}[t]
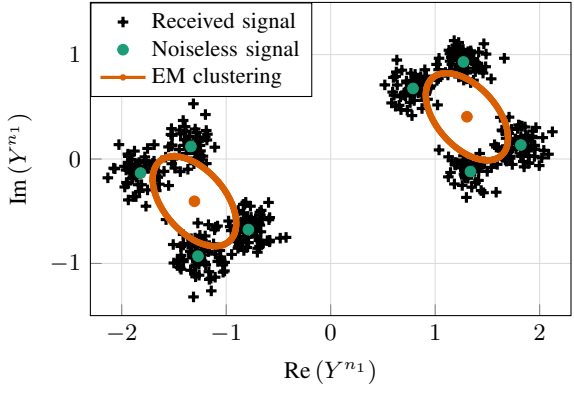

\begin{center}
\input{tikz/received_signal/double_column.tex}
\input{tikz/color_scheme.tex}                    
\pgfplotscreateplotcyclelist{colorbrew7dark}{%
rgb color={\PgfColorA}, mark=+, mark size = \MarkerSize, only marks, line width=\LineWidth\\%
rgb color={\PgfColorB}, mark=*, mark size = \MarkerSize, only marks, mark options = {fill=\colorB}\\%
rgb color={\PgfColorC}, mark=*, mark size = 0.3 * \MarkerSize, line width=\LineWidth\\%
rgb color={\PgfColorC}, mark=*, mark size = 0.3 * \MarkerSize, line width=\LineWidth\\%
rgb color={\PgfColorC}, mark=*, mark size = \MarkerSize, only marks\\%
}

\begin{tikzpicture}[define rgb/.code={\definecolor{mycolor}{RGB}{#1}},
                    rgb color/.style={define rgb={#1},mycolor}]
\begin{axis}[
    width=\FigureWidth,
    unit vector ratio*=1 1 1, 
    xlabel={$\text{Re}\left(Y^{n_1}\right)$},
    ylabel={$\text{Im}\left(Y^{n_1}\right)$},
    ytick distance=1.0,
    xtick distance=1.0,
    xmin=-2.3, xmax=2.3,
    ymin=-1.5, ymax=1.5,
    legend cell align={left},
    legend style={at={\LegendPositioning}, anchor=north west, nodes={scale=\LegendScale, transform shape}},
    axis line style={latex-latex},
    label style={font=\small},
    tick label style={font=\small},
    grid=both,
    major grid style={line width=.2pt,draw=gray!30},
    tick align=inside,
    tickpos=left,
    legend columns = 1,
    cycle list name=colorbrew7dark,
]
\addplot table[x=RE,y=IM]{tikz/received_signal/data/rx_scatter.dat};
\addplot coordinates {
(-0.7877, -0.6756)
(-1.3375, 0.1201)
(1.8201, 0.1343)
(1.2703, 0.9300)
(-1.2703, -0.9300)
(-1.8201, -0.1343)
(1.3375, -0.1201)
(0.7877, 0.6756)
};

\addplot table[x=RE,y=IM]{tikz/received_signal/data/em1.dat};
\addplot table[x=RE,y=IM]{tikz/received_signal/data/em2.dat};
\addplot coordinates {
(1.3039, 0.4049)
(-1.3039, -0.4049)
};

\legend{
Received signal,
Noiseless signal,
EM clustering
}
\end{axis}
\end{tikzpicture}
\caption {Example of received signal in case of $r=3$ users and SNR equals to $15$ dB. $8$ clusters can be distinctly seen at this SNR. Orange line shows the two component GM representation. Ellipse is 1-$\sigma$ contour of corresponding GM component covariance matrix. Noiseless signal $\pm H_1\pm H_2 \pm H_3$ is represented by green dots.\label{fig:received_signal}}
\end{center}
\end{figure}
To apply a TIN decoder we first construct the coarse channel estimate for the codeword with the highest received power. Recall that the fading coefficients corresponding to transmitted codewords are $H_1, \ldots, H_{r}$, $|H_1| \geq |H_2| \geq \ldots \geq |H_r|$. Given noiseless conditions and the BPSK modulation (which we use in our scheme), one can observe up to $2^{r}$ complex values within $n_1$ channel uses (see~\Fig{fig:received_signal}). The CSI estimation problem can be solved easily in this case.

In the noisy case, the problem is much more difficult, so we are going to estimate CSI for the strongest use only. To solve the problem we use clustering methods. Under BPSK modulation the received signal in every channel use can be clustered into two components ($\pm H \sqrt{P} + \tilde{Z}$). The clustering is performed in two-dimensional space corresponding to real and imaginary components of the received signal $Y^{n_1}$. This procedure tries to extract the signal corresponding to the strongest user and treat all other transmissions as a noise. Another benefit of this procedure is noise plus interference (NI) power estimation:  with the EM clustering, the covariance matrix of each component provides the estimate of the NI power, see~\Fig{fig:received_signal}. Under the block-fading channel model, this approach does not require any preambles or pilot symbols.

Thus, at the $i$-th TIN-SIC step we have the following approximate a posterior pdf for $H_i$ in the form of Gaussian mixture (GM)  
\begin{equation}
\label{eq:demodulator}
p(H_i) = \sum_{l = 1}^2 \mu^l \cn\left(\hat{H}_i^l, \hat{P}^l\right).
\end{equation}
After the clustering procedure, a two-component GM is representing the CSI pdf~\eqref{eq:demodulator}. Further decoding procedure must take into account both components, and we perform two decoding attempts with at most one being successful. The decoder is a polar list successive cancellation decoder with the cyclic redundancy check (CRC).

We calculate the decoder input log-likelihood ratios $l^{n_1}$ as follows
$$
\displaystyle
l^{n_1} = \log\left[\frac{\mathbb{E}_{h}p\left(Y^{n_1} | x = +\sqrt{P}, h\right)}{\mathbb{E}_{h}p\left(Y^{n_1} | x = -\sqrt{P}, h\right)}\right],
$$
where the expectation is taken over one of two GM components
$$
h\sim \cn\left(\hat{H}_i^l, \hat{P}^l \right),\quad l = 1,2.
$$

The list of codewords is a result of the decoding procedure. We first check the CRC and thus reduce the list size. The decoder outputs the most probable codeword from the remaining list or the empty set if the list is empty.
\begin{algorithm}
\caption{TIN decoder with on-the-fly CSI estimation\label{alg:TIN}}
\begin{algorithmic}
\State $Y'$ \Comment Residual signal after several SIC attempts
\State $p(H) \gets EM\left(Y'\right)$ \Comment perform the 2D EM clustering of the received signal.
\For {$l \in \{1, 2\}$} \Comment Try two hypotheses on $H$
\State Calculate $l^{n_1}$, $h\sim \cn\left(\hat{H}^l, \hat{P}^l \right)$ \Comment Demodulate given representation~\eqref{eq:demodulator}
\State $\hat{W} \gets decode\left(l^{n_1}\right)$ \Comment polar decoder
\If{$\hat{W} \neq \varnothing$}
\State return $\hat{W}$
\EndIf
\EndFor
\State return $\varnothing$
\end{algorithmic}
\end{algorithm}
The final TIN decoder is presented within Algorithm~\ref{alg:TIN}.
\subsection{Polar code construction}
In contrast to~\cite{Marshakov2019Polar} in this paper, all the users utilize the same polar code. Density evolution based polar code construction for Rayleigh fading channel was presented in~\cite{Trifonov2015}. In the case of TIN-SIC decoder, the equivalent users' channels become rather complex, so the methods of~\cite{Trifonov2015} cannot be used directly. At the same time, the codelength is small, so we used Monte-Carlo simulations in order to choose frozen subchannels. We constructed the code for $r=T$, i.e. for the largest collision order, we are going to resolve. During this simulation, the bit error for $i$-th bit has been evaluated given the $i-1$ previous bits. Given bit error distribution, the $k+c$ most reliable bits have been selected to be information bits, where $c$ is the CRC length.
\section{Numerical results and discussion}
\begin{figure}[t]
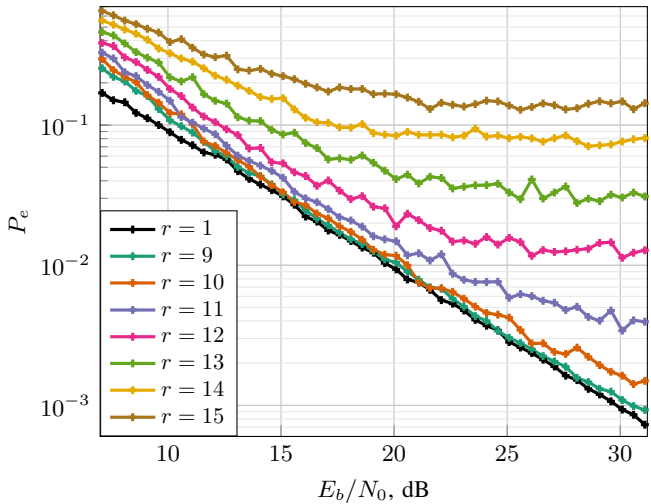

\begin{center}
\input{tikz/polar_pupe/double_column.tex}
\input{tikz/color_scheme.tex}                    
\pgfplotscreateplotcyclelist{colorbrew7dark}{%
rgb color={\PgfColorA}, mark=+, mark size = \MarkerSize, line width=\LineWidth\\%
rgb color={\PgfColorB}, mark=+, mark size = \MarkerSize, line width=\LineWidth\\%
rgb color={\PgfColorC}, mark=+, mark size = \MarkerSize, line width=\LineWidth\\%
rgb color={\PgfColorD}, mark=+, mark size = \MarkerSize, line width=\LineWidth\\%
rgb color={\PgfColorE}, mark=+, mark size = \MarkerSize, line width=\LineWidth\\%
rgb color={\PgfColorF}, mark=+, mark size = \MarkerSize, line width=\LineWidth\\%
rgb color={\PgfColorG}, mark=+, mark size = \MarkerSize, line width=\LineWidth\\%
rgb color={\PgfColorH}, mark=+, mark size = \MarkerSize, line width=\LineWidth\\%
}

\begin{tikzpicture}[define rgb/.code={\definecolor{mycolor}{RGB}{#1}},
                    rgb color/.style={define rgb={#1},mycolor}]

\begin{axis}[
    width=\FigureWidth, height=\FigureHeight,
    ymode=log,
    xlabel={${E_b}/{N_0}$, dB},
    ylabel={$P_e$},
    xmin=7.0,    xmax=31.2,
    ymin=6e-4, ymax=0.7,
    label style={font=\small},
    tick label style={font=\small},
    ylabel style={yshift=-0.8mm},
    legend cell align={left},
    legend style={at={\LegendPositioning}, anchor=south west, nodes={scale=\LegendScale, transform shape}},
    axis line style={latex-latex},
    grid=both,
    grid style={line width=.1pt, draw=gray!20},
    major grid style={line width=.2pt,draw=gray!50},
    tick align=inside,
    tickpos=left,
    legend columns=1,
    cycle list name=colorbrew7dark,
]

\addplot table[x=EBNO,y=FER]{tikz/polar_pupe/data/K01.dat};
\addplot table[x=EBNO,y=FER]{tikz/polar_pupe/data/K09.dat};
\addplot table[x=EBNO,y=FER]{tikz/polar_pupe/data/K10.dat};
\addplot table[x=EBNO,y=FER]{tikz/polar_pupe/data/K11.dat};
\addplot table[x=EBNO,y=FER]{tikz/polar_pupe/data/K12.dat};
\addplot table[x=EBNO,y=FER]{tikz/polar_pupe/data/K13.dat};
\addplot table[x=EBNO,y=FER]{tikz/polar_pupe/data/K14.dat};
\addplot table[x=EBNO,y=FER]{tikz/polar_pupe/data/K15.dat};

\legend{
$r  = 1$,
$r  = 9$,
$r  = 10$,
$r  = 11$,
$r  = 12$,
$r  = 13$,
$r  = 14$,
$r  = 15$
}

\end{axis}
\end{tikzpicture}
{\vskip -0.3cm}
\caption {TIN-SIC PUPE. Simulation results for a slot decoding performance for different $r$. Polar code with $k=100$ information bits, $n_1=512$ and BPSK modulation. List size 64, CRC-21 (polar code with equivalent $\tilde{k}=k+c=121$ bits).\label{fig:polar_pupe}}
\end{center}
\end{figure}

Let us start with the slot decoding performance. In comparison with the previous work~\cite{FadingFull2019}, we have selected a slightly longer slot ($n_1 = 512$ channel uses) and utilize polar slot codes. We use a successive cancellation list decoder with the list size equals to $64$. Recall that $r$ denotes the number of users transmitting in a slot. The PUPE ($P_e$) has been evaluated for $r$ up to $15$. The most surprising result is that the TIN-SIC scheme can resolve a relatively high number of simultaneous transmissions (see~\Fig{fig:polar_pupe}). For $r \leq 9$, the PUPE performance does not significantly deviate from the $r = 1$ performance curve. $P_e$ remains below the target $P_e\leq \varepsilon = 0.1$ up to $r=14$. The LDPC based solution from~\cite{FadingASYNC2019} has achieved this result for $r \leq 8$. Given the fact that the weakest users in a slot deliver the most fraction of $P_e$, we concluded to use $T=14$ in our energy efficiency curve.
We have tested CRC length equal to $11$, $16$ and $21$. Finally, the CRC length has been selected to be equal to $21$ bits to achieve sufficiently low false alarm rate ($<10^{-2}P_e$) while performing a high number of decoding attempts (especially for $r<T$).

\begin{figure}[t]
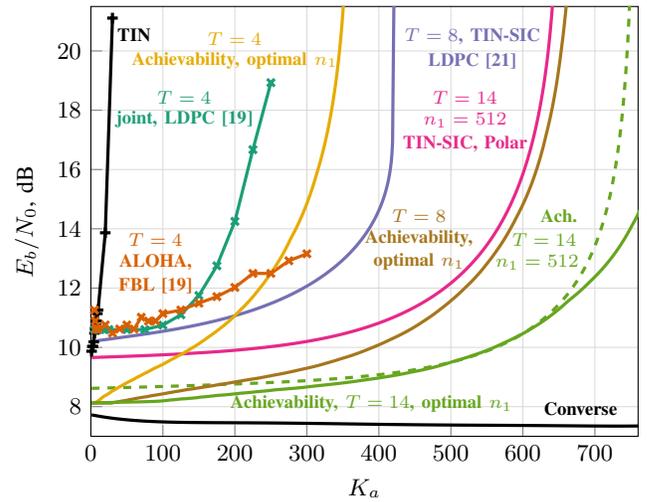

\begin{center}
\input{tikz/ebno_ka/double_column.tex}
\input{tikz/color_scheme.tex}                    
\pgfplotscreateplotcyclelist{colorbrew7dark}{%
rgb color={\PgfColorA}, mark=+, mark size = \MarkerSize, line width=\LineWidth\\%
rgb color={\PgfColorB}, mark=x, mark size = \MarkerSize, line width=\LineWidth\\%
rgb color={\PgfColorC}, mark=x, mark size = \MarkerSize, line width=\LineWidth\\%
rgb color={\PgfColorD}, mark=none, line width=\LineWidth\\%
rgb color={\PgfColorE}, mark=none, line width=\LineWidth\\%
rgb color={\PgfColorF}, mark=none, dashed, line width=\LineWidth\\%
rgb color={\PgfColorG}, mark=none, line width=\LineWidth\\%
rgb color={\PgfColorH}, mark=none, line width=\LineWidth\\%
rgb color={\PgfColorF}, mark=none, line width=\LineWidth\\%
rgb color={\PgfColorA}, mark=none, line width=\LineWidth\\%
}

\begin{tikzpicture}[define rgb/.code={\definecolor{mycolor}{RGB}{#1}},
                    rgb color/.style={define rgb={#1},mycolor}]
\begin{axis}[
    width=\FigureWidth, height=\FigureHeight,
    xlabel={$K_a$},
    ylabel={${E_b}/{N_0}$, dB},
    ytick distance=2,
    xtick distance=100,
    xmin=0,    xmax=760,
    ymin=7, ymax=21.5,
    legend cell align={left},
    legend style={at={\LegendPositioning}, anchor=north west, nodes={scale=\LegendScale, transform shape}},
    axis line style={latex-latex},
    label style={font=\small},
    tick label style={font=\small},
    grid=both,
    major grid style={line width=.2pt,draw=gray!30},
    tick align=inside,
    tickpos=left,
    legend columns = 1,
    cycle list name=colorbrew7dark,
]
\input{tikz/color_scheme.tex}
\addplot table[x=Ka,y=EBNO]{tikz/ebno_ka/data/tin.dat};
\addplot table[x=Ka,y=EBNO]{tikz/ebno_ka/data/aloha4_ldpc.dat};
\addplot table[x=KA,y=EBNO]{tikz/ebno_ka/data/aloha4_fbl.dat};
\addplot table[x=Ka,y=EBNO]{tikz/ebno_ka/data/ldpc_T8.dat};
\addplot table[x=Ka,y=EBNO]{tikz/ebno_ka/data/polar_T14.dat};
\addplot table[x=Ka,y=EBNO]{tikz/ebno_ka/data/achievability_k100_n512_T14.dat};
\addplot table[x=Ka,y=EBNO]{tikz/ebno_ka/data/achievability_envelope_T4.dat};
\addplot table[x=Ka,y=EBNO]{tikz/ebno_ka/data/achievability_envelope_T8.dat};
\addplot table[x=Ka,y=EBNO]{tikz/ebno_ka/data/achievability_envelope_T14.dat};
\addplot table[x=Ka,y=EBNO]{tikz/ebno_ka/data/converse.dat};
\node[above,\colorA] at (60,20) {\scriptsize{\textbf{TIN}}};

\node[above,\colorB] at (130,17.9) {\scriptsize{\textbf{$T= 4$}}};
\node[above,\colorB] at (130,17.1) {\scriptsize{\textbf{joint, LDPC~\cite{FadingISIT2019}}}};

\node[above,\colorC] at (90,13.1) {\scriptsize{\textbf{$T= 4$}}};
\node[above,\colorC] at (90,12.3) {\scriptsize{\textbf{ALOHA,}}};
\node[above,\colorC] at (90,11.5) {\scriptsize{\textbf{FBL~\cite{FadingISIT2019}}}};

\node[above,\colorD] at (530, 19.9) {\scriptsize{\textbf{$T=8$, TIN-SIC}}};
\node[above,\colorD] at (530, 19.1) {\scriptsize{\textbf{LDPC~\cite{FadingASYNC2019}}}};

\node[above,\colorE] at (520, 17.9) {\scriptsize{\textbf{$T=14$}}};
\node[above,\colorE] at (520, 17.1) {\scriptsize{\textbf{$n_1 = 512$}}};
\node[above,\colorE] at (520, 16.3) {\scriptsize{\textbf{TIN-SIC, Polar}}};

\node[above,\colorF] at (650, 13.9) {\scriptsize{\textbf{Ach.}}};
\node[above,\colorF] at (630, 13.1) {\scriptsize{\textbf{$T=14$}}};
\node[above,\colorF] at (620, 12.3) {\scriptsize{\textbf{$n_1 = 512$}}};

\node[above,\colorF] at (390, 7.5) {\scriptsize{\textbf{Achievability, $T=14$, optimal $n_1$}}};

\node[above,\colorG] at (200, 19.9) {\scriptsize{\textbf{$T=4$}}};
\node[above,\colorG] at (200, 19.1) {\scriptsize{\textbf{Achievability, optimal $n_1$}}};

\node[above,\colorH] at (456, 13.9) {\scriptsize{\textbf{$T=8$}}};
\node[above,\colorH] at (456, 13.1) {\scriptsize{\textbf{Achievability,}}};
\node[above,\colorH] at (456, 12.3) {\scriptsize{\textbf{optimal $n_1$}}};

\node[above,\colorA] at (680, 7.4) {\scriptsize{\textbf{Converse}}};

\end{axis}
\end{tikzpicture}
{\vskip -0.3cm}
\caption {$K_a$ vs $E_b/N_0$ for $\varepsilon=0.1$, $n=30000$, $k=100$ information bits for different schemes (LDPC from~\cite{FadingISIT2019} and polar) with different $T$. Achievability bounds for fixed code length ($n_1 = 512$, dashed line) and for adjustable code length (optimal $n_1$) shown for reference as well as converse bound~\cite{Polyanskiy2013Fading}. ALOHA plus finite blockelngth (FBL) taken from~\cite{FadingISIT2019}.\label{fig:ebno_ka}}
\end{center}
\end{figure}

The energy efficiency is the minimum energy required to serve $K_a$ users in a frame of length $n=30000$ with PUPE less than $\varepsilon = 0.1$ in our reference setup (see~\Fig{fig:ebno_ka}). In comparison with LDPC codes, one can observe better performance when the system load is low due to better polar codes performance. At high system load, the performance is still better due to the TIN-SIC ability to extract a high number of simultaneous transmission from every slot. The optimal over $n_1$ achievability bound means the minimum required energy among all possible slot lengths $n_1$. For example, the green dashed line represents the achievability bound for $n_1 = 512$. The solid green line represents the optimal over $n_1$ achievability bound. Both lines touch each other at $K_a\approx 600$. Thus, given $K_a < 600$ the longer slot length is optimal, while shorter codes perform better at $K_a > 600$. The same achievability for $T=4$ has been constructed (yellow line). On the other hand, the achievability bound from~\cite{FadingISIT2019} (ALOHA plus FBL) provides better results at high loads.

One can see that the polar code curve passes about $\Delta<1$ dB below the achievability bound for $n_1 = 512$ at low loads (dashed green versus purple lines,~\Fig{fig:ebno_ka}. This result corresponds to $k=100$ information bits and CRC length $c = 21$. A more careful look at this gap allows us to formulate the main result of our practical scheme: the performance loss caused by the additional CRC bits, $\Delta \approx 10 \cdot \log_{10}{\left(1 +c/k\right)}$. Thus, the additional CRC bits is the major source of the energy efficiency loss with respect to achievability bound.

To the best knowledge of the authors, the derived achievability bound and the proposed polar code based scheme outperform the existing results from the literature on the unsourced random access in the fading MAC. Further research should be devoted to the following directions: (a) developing an achievability bound with non-ideal SIC step taken into account, (b) improving the decoding algorithm for polar code based scheme, and (c) considering multiple antennas at the receiver.
\bibliographystyle{IEEEtran}
\bibliography{main}
\end{document}